\documentclass[10pt,twocolumn,floatfix,superscriptaddress,preprintnumbers,amsmath,amssymb,prb]{revtex4-2}
\usepackage{graphicx}
\usepackage{dcolumn}
\usepackage{bm}
\usepackage{amssymb,amsmath}
\usepackage{float}
\usepackage{color}
\usepackage{ulem}

\begin{document}

\preprint{preprint(\today)}

\title{Time reversal invariant single gap superconductivity with upper critical field larger than Pauli limit in NbIr$_2$B$_2$ }

\author{Debarchan Das}
\email{debarchandas.phy@gmail.com}
\affiliation{Laboratory for Muon Spin Spectroscopy, Paul Scherrer Institute, CH-5232 Villigen PSI, Switzerland}
\author{Karolina G\'{o}rnicka}
\affiliation{Faculty of Applied Physics and Mathematics, Gda\'{n}sk University of Technology, ul. Narutowicza 11/12, Gda\'{n}sk 80–233, Poland}
\affiliation{Advanced Materials Centre, Gda\'{n}sk University of Technology, ul. Narutowicza 11/12, Gda\'{n}sk 80–233, Poland}
\author{Zurab Guguchia}
\affiliation{Laboratory for Muon Spin Spectroscopy, Paul Scherrer Institute, CH-5232 Villigen PSI, Switzerland}
\author{Jan Jaroszynski}
\affiliation{National High Magnetic Field Laboratory, Florida State University, Tallahassee FL 32310, USA}
\author{Robert J. Cava}
\affiliation{Department of Chemistry, Princeton University, Princeton, NJ 08540, USA}
\author{Weiwei Xie}
\affiliation{Department of Chemistry and Chemical Biology, Rutgers University, Piscataway, NJ 08854, USA}
\author{Hubertus Luetkens}
\affiliation{Laboratory for Muon Spin Spectroscopy, Paul Scherrer Institute, CH-5232 Villigen PSI, Switzerland}
\author{Tomasz Klimczuk}
\email{tomasz.klimczuk@pg.edu.pl}
\affiliation{Faculty of Applied Physics and Mathematics, Gda\'{n}sk University of Technology, ul. Narutowicza 11/12, Gda\'{n}sk 80–233, Poland}
\affiliation{Advanced Materials Centre, Gda\'{n}sk University of Technology, ul. Narutowicza 11/12, Gda\'{n}sk 80–233, Poland}

\begin{abstract}

Recently, compounds with noncentrosymmetric crystal structure have attracted much attention for providing a rich playground in search for unconventional superconductivity. NbIr$_2$B$_2$ is a new member to this class of materials harboring superconductivity below $T_{\rm c} = 7.3(2)$~K and very high upper critical field that exceeds Pauli limit. Here we report on muon spin rotation ($\mu$SR) experiments probing the temperature and field dependence of effective magnetic penetration depth in this compound. Our transverse-field -$\mu$SR results suggest a fully gaped $s$-wave superconductvity. Further, the estimated high value of upper critical field is also supplemented by high field transport measurements.  Remarkably, the ratio $T_{\rm c}$/$\lambda^{-2}(0)$ obtained for NbIr$_2$B$_2$ ($\sim$2)  is comparable to those of unconventional superconductors. Zero-field $\mu$SR data reveals no significant change in the muon spin relaxation rate above and below $T_{\rm c}$, evincing that time-reversal symmetry is preserved in the superconducting state. The presented results will stimulate theoretical investigations to obtain a microscopic understanding of the origin of superconductivity with preserved time reversal symmetry in this unique noncentrosymmetric system.

\end{abstract}


\maketitle

\section{Introduction}

 The crystal structure of a non-centrosymmetric superconductor (NCS) lacks a centre of inversion favoring an electronic antisymmetric spin orbit coupling (ASOC) to subsist by symmetry\cite{Bauerbook, Smidmanreview,Yip, Ghosh, Tian}. For sufficiently large ASOC, one may expect mixing of spin-singlet and spin-triplet Copper pairing channels leading to many exotic superconducting properties namely, very high upper critical fields higher than the Pauli limit, nodes in the superconducting gaps, appearance of tiny spontaneous fields below the superconducting transition temperature, $T_c$, breaking time-reversal symmetry (TRS) etc\cite{Ghosh, Tian}. For this reason, NCSs are of significant interest and have become an intensively studied topic in contemporary condensed matter research.

 There are quite a few examples of NCSs in the literature e.g. CePt$_3$Si\cite{BauerCePtSi}, Ce(Rh,Ir)Si$_3$\cite{Kimura, Sugitani}, LaNiC$_2$\cite{Hillier}, La$_7$Ir$_3$\cite{Barker}, Re$_x$T$_y$ series (T = 3$d$--5$d$ early transition metals)\cite{RPSingh, DSingh,ShangRebased,Shangnpj}, Mg$_{10}$Ir$_{19}$B$_{16}$\cite{Aczel}, ARh$_2$B$_2$(A= Nb and Ta)\cite{Carnicom, Mayoh1, Mayoh2}   etc \cite{ Ghosh, Tian,DasLaPtSi,KhasanovBeAu, DasZrRuAs,Shangnjp, Sajilesh1, Sajilesh2, Arushi} . Amongst these wide range of materials, compounds without any magnetic $f$ electron element are of particular interest because it allows to study the intrinsic pairing mechanisms in NCSs. Thus, the search for new NCSs of this class has become an exigent goal. Very recently, we have discovered two novel Ir--based NCSs NbIr$_2$B$_2$ and TaIr$_2$B$_2$ forming a unique low symmetry $Cc$ noncentrosymmetric crystal structure\cite{Gornicka}.  First-principles calculations and symmetry analysis suggest that these materials are topological Weyl metals in the normal state\cite{Gao}. Bulk measurements reveal superconducting properties having $T_c$'s 7.2 and 5.1~K along with considerably high value of the upper critical fields 16.3 and 14.7~T respectively. Interestingly, theoretical calculations signals a possible multigap scenario for NbIr$_2$B$_2$. Therefore, detailed microscopic experimental study is essential to address this intriguing aspect. Muon spin rotation/relaxation ($\mu$SR)\cite{Sonier,Blundellbook, Hillierreview} is a very sensitive technique to probe the superconducting gaps and the nature of the pairing in superconductors. In case of a type-II superconductor, the mixed or vortex state creates flux line lattice (FLL) which gives rise to an inhomogeneous spatial distribution of local magnetic fields influencing the muon spin depolarization rate. It is directly related to the magnetic penetration depth $\lambda$ which is one of the fundamental length scales of a superconductor. The temperature dependence of $\lambda$ is sensitive to the structure of the superconducting gap. Moreover, zero-field $\mu$SR is a very powerful tool to detect the presence of infinitesimally small magnetic field which is crucial in verifying whether TRS is broken in the superconducting state.
\begin{figure*}[htb!]
\includegraphics[width=1.0\linewidth]{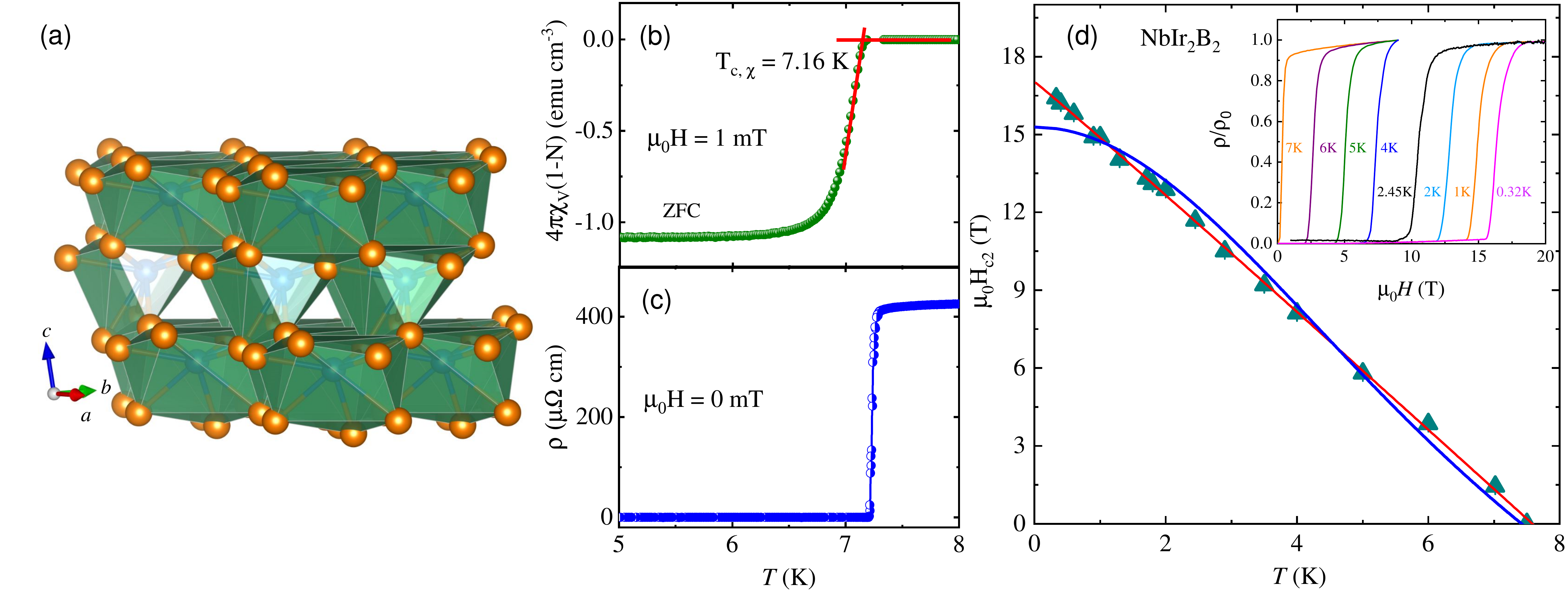}
\caption{(Color online)  (a) crystal structure of NbIr$_2$B$_2$. Orange spheres represent Nb atoms and Ir atoms are shown as blue spheres. For better clarity B-B dimers are not presented in the figure. (b) temperature dependence of magnetic susceptibility measured in zero field cooled condition under 1~mT applied field. (c) Zero field electrical resistivity ($\rho$) as a function of temperature. (d) $\mu_0H_{\rm c2} (T)$ phase diagram. Red and blue solid lines correspond to the fitting using different models as discussed in the text. Inset: field dependence of normalised resistivity ($\rho/\rho_0$) measured at different temperatures. Only few selected temperatures are shown for clarity.}
\label{fig1}
\end{figure*}

In this report, we present the results of our detailed $\mu$SR investigation performed on NbIr$_2$B$_2$ aiming to unravel the superconducting gap structure and to check whether the TRS is broken or preserved in the superconducting state. In addition, we also report electrical transport measurements under high magnetic fields (up to 20~T) down to 300~mK to extract more reliable value of upper critical field. Our results evince fully gap $s$-wave pairing with preserved time-reversal symmetry in NbIr$_2$B$_2$ possessing a very high value of upper critical field.

\section{EXPERIMENTAL DETAILS }
A polycrystalline sample of NbIr$_2$B$_2$ was prepared by a solid state reaction method of the constituent elements. The detailed procedure of sample preparation can be found in Ref.\cite{Gornicka}. Phase purity of the polycrystalline sample was checked by powder X-ray diffraction (XRD) using Cu-K$\alpha$ radiation and other metalographic experiments such as scanning electron microscopy (SEM) and energy-dispersive X-Ray spectroscopy (EDX). The details of Rietveld analysis together with crystallographic data can be found in Ref.\cite{Gornicka}. A NbIr$_2$B$_2$ sample for magneto-transport measurements was prepared in a bar form with four 50~$\mu$m dia platinum wire leads spark-welded to the sample surface. Transverse-field (TF) and zero-field (ZF) $\mu$SR experiments were carried out at the Paul Scherrer Institute (Villigen, Switzerland). The measurements down to 1.5~K were performed at $\pi$M3.2 beamline using GPS spectrometer and measurements down to 270~mK were conducted at $\pi$E1 beamline on DOLLY spectrometer. The powdered sample was pressed into a 7~mm pellet which was then mounted on a Cu holder using GE varnish. This holder assembly was then mounted in the respective spectrometer cryostats. Both spectrometers are equipped with a standard veto setup \cite{Amato} providing a low-background $\mu$SR signal. All the TF experiments were performed after field-cooled-cooling the sample. The $\mu$SR time spectra were analyzed using the MUSRFIT software package~\cite{Suter}.

\section{RESULTS}

\subsection{Crystal structure, sample characterization and high magnetic field measurements}
Figure~\ref{fig1}(a) shows the crystal structure of NbIr$_2$B$_2$.  The same type of polyhedrons, oriented in different directions, are formed by Nb atoms with Ir atom inside. For clarity B-B dimers, which are located in voids of the polyhedrons, are not presented in the figure. The superconductivity of the samples was confirmed by magnetic susceptibility and electrical resistivity measurements.  Figure~\ref{fig1}(b) shows the temperature dependence of magnetic susceptibility (measured in zero field cooled condition in an applied field of 1~mT) which manifests a diamagnetic signal below  $T_{\rm c, \chi}$ = 7.16~K concurrent with the offset of zero-resistivity (Fig~\ref{fig1}(c)). These results highlight the good quality of the sample with superconducting properties matching well with our previous report\cite{Gornicka}.

Figure~\ref{fig1}(d) represents the upper critical field-temperature [$\mu_0H_{\rm c2}(T)$] phase diagram determined from field dependent electrical transport measurements performed at National High Magnetic Field Laboratory (FSU). The data points were obtained from $\rho$ vs. $\mu_0H$ measurements carried out at constant temperatures. Selected curves are shown in the inset of Figure~\ref{fig1}(d) (presented as normalised resistivity). The critical field at each temperature was estimated as a midpoint of the transition. It is worth noting that the superconducting transition only slightly broadens under the highest applied magnetic field. It is quite evident from the linear temperature  dependence of upper critical field, that $\mu_0H_{\rm c2}(T)$ can not be modeled using the Werthamer-Helfand-Hohenberg (WHH) model \cite{Werthamer} which accounts for Pauli limiting and spin-orbit scattering effects. We used  the following model to describe the temperature dependence of upper critical field (red solid line), $\mu_0H_{\rm c2} (T) = \mu_0H_{\rm c2} (0)[1-(T/T_{\rm c})^n]$ yielding $n$ = 1.02(2) and $\mu_0H_{\rm c2} (0)$ = 17.0(1)~T. Notably, thus obtained value of $\mu_0H_{\rm c2} (0)$ is considerably higher than the Pauli-limiting field [13.3(1)~(T)] signaling non-BCS type superconductivity in NbIr$_2$B$_2$. To model the data, we also tested Ginzburg–Landau expression\cite{Tinkham}:

\begin{equation}
\mu_0H_{c2}(T)=\mu_0H_{c2}(0)\frac{(1-t^2)}{(1+t^2)}
\label{GL}
\end{equation}
\noindent where $t = T/T_{\rm c}$ and $T_{\rm c}$ is the transition temperature at zero magnetic field. As seen from the fitting (blue solid line), this model does not work in the present case.

\begin{figure*}
\includegraphics[width=0.9\linewidth]{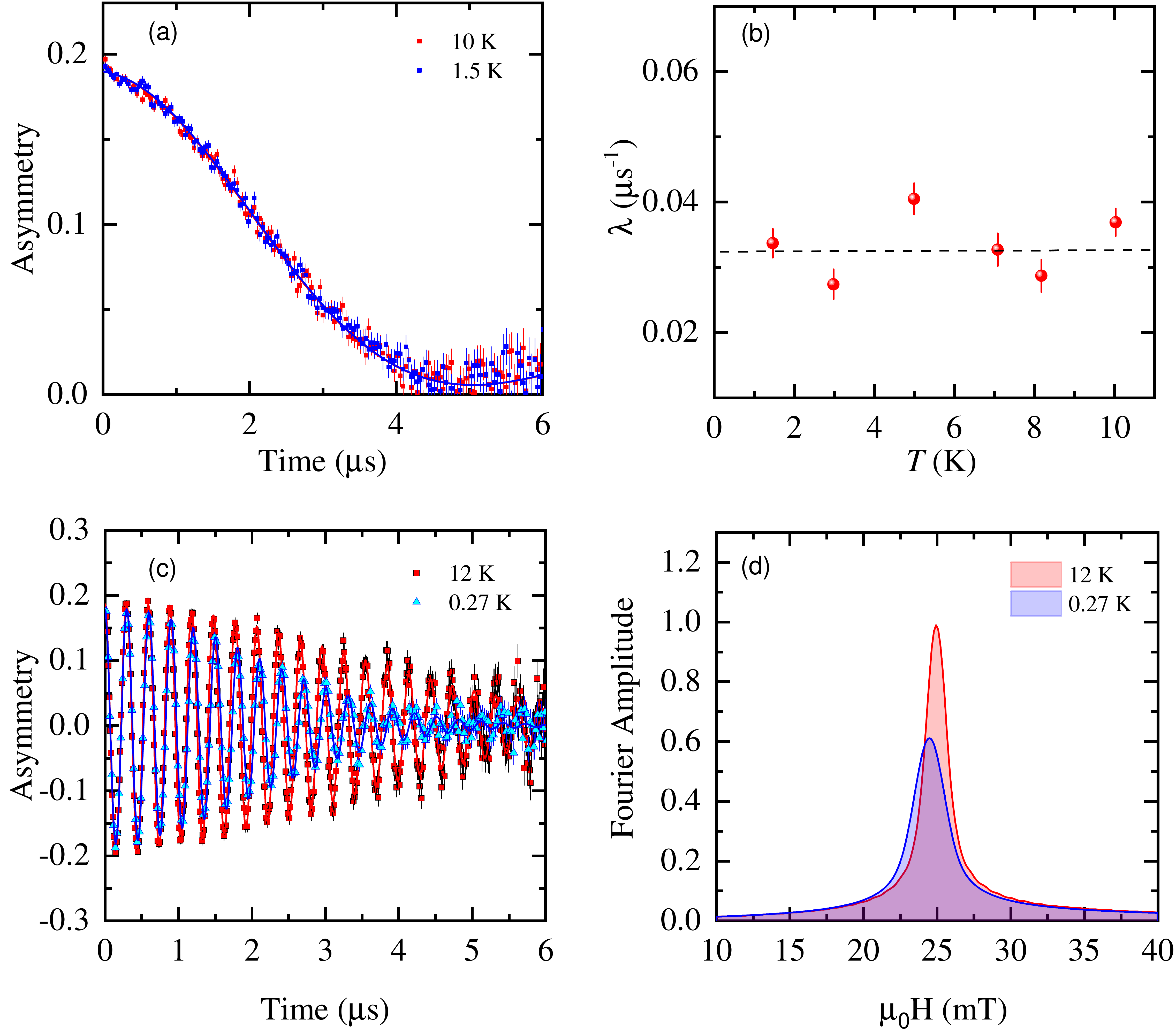}
\caption{(Color online) (a) ZF $\mu$SR asymmetry spectra recorded at 1.5 and 10~K for NbIr$_2$B$_2$. (b) Temperature dependence of the electronic relaxation rate measured in zero magnetic field. (c) Transverse-field (TF) ${\mu}$SR time spectra obtained above and below $T_{\rm c}$ for \mbox{NbIr$_2$B$_2$} in an applied field of 25~mT (after field cooling the sample from above $T_{\rm c}$). (d) Fourier spectra at 0.27~K (blue) and 12~K (red)  obtained by fast Fourier transformation of the ${\mu}$SR time spectra from panel (c).}
\label{fig2}
\end{figure*}

\begin{figure}[htb!]
\includegraphics[width=0.9\linewidth]{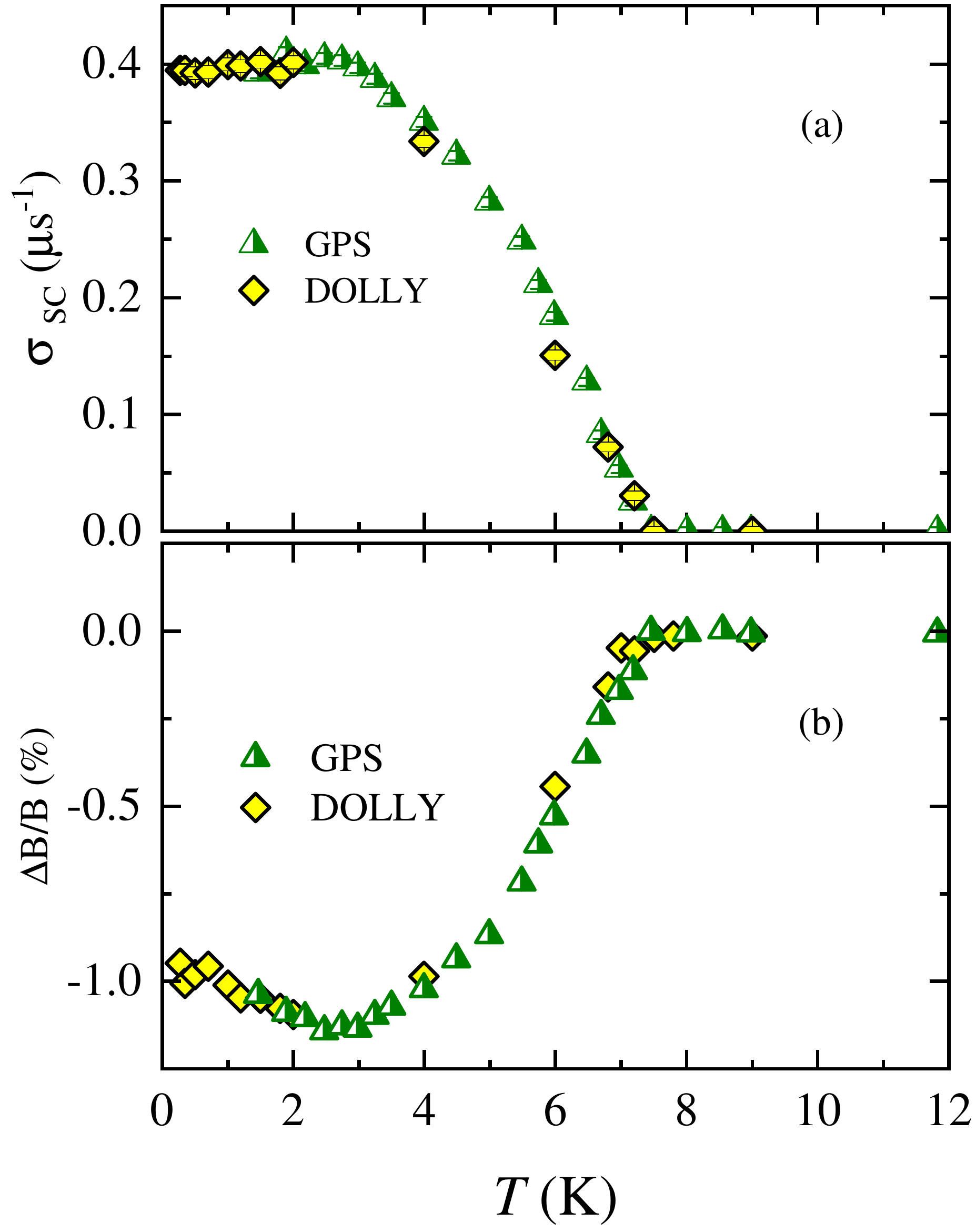}
\caption{(Color online)  a) Temperature evolution of the superconducting muon spin depolarization rate ${\sigma}_{\rm sc}$ of NbIr$_2$B$_2$ measured in an applied magnetic field of 25~mT. (b) Temperature dependence of the relative change of the internal field normalized to the external applied field, $\Delta B/B_{\rm ext}\left(= \frac {B_{\rm int}-B_{\rm ext}}{B_{\rm ext}}\right)$.}
\label{fig3}
\end{figure}

\subsection{ZF-${\mu}$SR Measurements}

In NCSs, due to the admixture of spin-singlet and spin-triplet superconducting channels, small magnetic moments
associated with the formation of spin-triplet electron pairs might appear in the superconducting state. Thus, NCSs are prime candidates to search for novel superconductors showing TRS breaking. To explore this tempting issue in this novel NCS NbIr$_2$B$_2$, we first performed ZF-${\mu}$SR experiments above and below $T_{{\rm c}}$ to detect any possible spontaneous magnetic fields which leads to broken TRS. Figure~\ref{fig2}a shows the ZF-${\mu}$SR asymmetry spectra for temperatures above and below $T_{{\rm c}}$. We do not observe any noticeable difference in the spectra suggesting absence of any spontaneous field in the superconducting state. The ZF-${\mu}$SR asymmetry spectra can be well described by a damped Gaussian Kubo-Toyabe depolarization function\cite{Toyabe} $A_{\rm ZF}(t) = A_0~G_{KT}~\exp(-\Lambda t)$ where, $A_0$ is the initial asymmetry, $G_{KT}$ is Gaussian Kubo-Toyabe (KT) function\cite{Toyabe} which accounts for an isotropic Gaussian distribution of randomly oriented static (or quasistatic) local fields at the muon sites and $\Lambda$ is the electronic relaxation rate. Figure~\ref{fig2}b represents the temperature dependence of $\Lambda$ which shows no considerable enhancement across $T_{\rm c}$. The maximum possible spontaneous flux density due to superconductivity can be estimated using \mbox{($\Lambda|_{1.5~\rm K}-\Lambda|_{10~\rm K})/(2\pi\gamma_{\mu}) = 0.43~\mu$T} which is several times smaller than that seen for well known TRS breaking superconductors \cite{Hillier, Barker,ShangRebased, LukeTRS}. We note that even though the field resolution of the instrument is finite,  it is sufficient to detect an internal field of the magnitude found in other superconductors where time reversal symmetry breaking was observed. Therefore, it can be concluded that the time-reversal symmetry is preserved in the superconducting state of NbIr$_2$B$_2$. Electronic structure calculation presented in our earlier report \cite{Gornicka}, manifests that in spin-split bands, $E(k) = E(-k)$ degeneracy is kept while spin direction is flipped without giving rise to any net moment as $k$ is changed to -$k$ . Therefore, absence of time reversal symmetry breaking in the superconducting state is consistent with the electronic structure calculation.

\subsection{ TF-${\mu}$SR  Measurements}

Figure~\ref{fig2}c represents TF-$\mu$SR spectra for NbIr$_2$B$_2$ measured in an applied magnetic field of ~25~mT at temperatures above (12~K) and below (0.27~K) $T_{\rm c}$. Above $T_{\rm c}$, we observed a small relaxation in TF-$\mu$SR spectra  due to the presence of random local fields associated with the nuclear magnetic moments. However, in the superconducting state, the formation of FLL causes an inhomogeneous distribution of magnetic field which increases the relaxation rate of the $\mu$SR signal. Assuming a Gaussian field distribution, we analyzed the observed TF-$\mu$SR asymmetry spectra using the following functional form
\begin{equation}
A_{\rm TF}(t)=A_{0}\exp\left(\sigma^2t^2/2\right)\cos\left(\gamma_{\mu}B_{\rm int}t+\varphi\right)
\label{ATF}
\end{equation}
\noindent where $A_{0}$ refers to the initial asymmetry, $\gamma_\mu/(2{\pi})\simeq 135.5$~MHz/T is the muon gyromagnetic ratio, and ${\varphi}$ is the initial phase of the muon-spin ensemble, $B_{\rm int}$ corresponds to the internal magnetic field at the muon site, respectively and $\sigma$ is the total relaxation rate. Here, $\sigma$ is related to the superconducting relaxation rate, $\sigma_{\rm SC}$, following the relation $\sigma=\sqrt{\sigma_{\rm nm}^2+\sigma_{\rm SC}^2}$ where $\sigma_{\rm nm}$ is the nuclear contributions which is assumed to be temperature independent. For estimating $\sigma_{\rm SC}$, we considered the value of $\sigma_{\rm nm}$ obtained above $T_{\rm c}$ where only nuclear magnetic moments contribute to the muon depolarization rate $\sigma$ and kept it fixed. The fits to the observed spectra with Eq.~\ref{ATF} are shown in solid lines in Fig.~\ref{fig2}c. Figure~\ref{fig2}d depicts the Fourier transform amplitudes of the TF-${\mu}$SR asymmetry spectra recorded at 12~K and 0.27~K (Fig.~\ref{fig2}c). We observed a sharp peak in the Fourier amplitude around 25~mT (external applied field) at 12~K confirming homogeneous field distribution throughout the sample. Notably, a fairly broad signal with a peak position slightly shifted to lower value (diamagnetic shift) was seen at 0.27~K evincing the fact that the sample is indeed in the superconducting mixed state where the formation of the FLL causes such broadening of the line shape.

In Figure~\ref{fig3}a, we have presented ${\sigma}_{\rm sc}$ as a function of temperature for NbIr$_2$B$_2$ measured at an applied field of 25~mT. Below $T_{\rm c}$, the relaxation rate ${\sigma}_{\rm sc}$ increases from zero due to inhomogeneous field distribution caused by the formation of FLL, and saturates at low temperatures. In the following section, we show that the observed temperature dependence of ${\sigma}_{\rm sc}$, which reflects the topology of the superconducting gap, is consistent with the presence of the single gap on the Fermi surface of NbIr$_2$B$_2$. Figure~\ref{fig3}b shows the temperature dependence of the relative change of the internal field normalized to the external applied field, $\Delta B/B_{\rm ext}\left(= \frac {B_{\rm \rm int}-B_{\rm ext}}{B_{\rm ext}}\right)$. As seen from the figure, internal field values in the superconducting state ($i.e.~T<T_{\rm c}$) are lower than the applied field because of the diamagnetic shift, expected for type-II superconductors.

Considering a perfect triangular vortex lattice, the muon spin depolarization rate ${\sigma}_{\rm sc}(T)$ is directly related to the London magnetic penetration depth ${\lambda}(T)$ by~\cite{Brandt,Brandt2}:
\begin{equation}
\frac{\sigma_{\rm sc}^2(T)}{\gamma_\mu^2}=0.00371\frac{\Phi_0^2}{\lambda^4(T)}.
\label{Sigma}
\end{equation}
\noindent Here, $\Phi_0$ = 2.068~$\times$~10$^{-15}$~Wb is the magnetic flux quantum. We note that Eq.~\ref{Sigma} is only valid when the separation between the vortices is smaller than ${\lambda}$ which is presumed to be field independent in this model~\cite{Brandt}. To gain insight about the superconducting gap structure of NbIr$_2$B$_2$ and estimate various parameters defining the superconducting state of this system, we analyzed the temperature dependence of the magnetic penetration depth, ${\lambda}(T)$.

Within the London approximation ($\lambda \gg {\xi}$), ${\lambda}(T)$ can be described by the following expression,~\cite{Suter,Tinkham, Prozorov}
\begin{equation}
\frac{\lambda^{-2}(T,\Delta_{0,i})}{\lambda^{-2}(0,\Delta_{0,i})}=
1+\frac{1}{\pi}\int_{0}^{2\pi}\int_{\Delta(_{T,\varphi})}^{\infty}\left(\frac{\partial f}{\partial E}\right)\frac{EdE}{\sqrt{E^2-\Delta_i(T,\varphi)^2}},
\label{Lambda}
\end{equation}
\noindent where $f=\left[1+\exp\left(E/k_{\rm B}T\right)\right]^{-1}$ is the Fermi function, $\varphi$ is the azimuthal angle  along the Fermi surface and ${\Delta}_{0,i}\left(T\right)={\Delta}_{0,i}~{\Gamma}\left(T/T_{\rm c}\right)$~g($\varphi$). ${\Delta}_{0,i}$ is the maximum gap value at $T=0$~K. The temperature dependence of the gap is described by the expression \mbox {${\Gamma}\left(T/T_{\rm c}\right)=\tanh\left\{1.82\left[1.018\left(T_{\rm c}/T-1\right)\right]^{0.51}\right\}$}~\cite{carrington}. The angular dependence g($\varphi$) takes the value 1 for $s$-wave gap whereas for a nodal $d$-wave gap it is $\mid cos(2\varphi)\mid$.

\begin{figure}[htb!]
\includegraphics[width=0.9\linewidth]{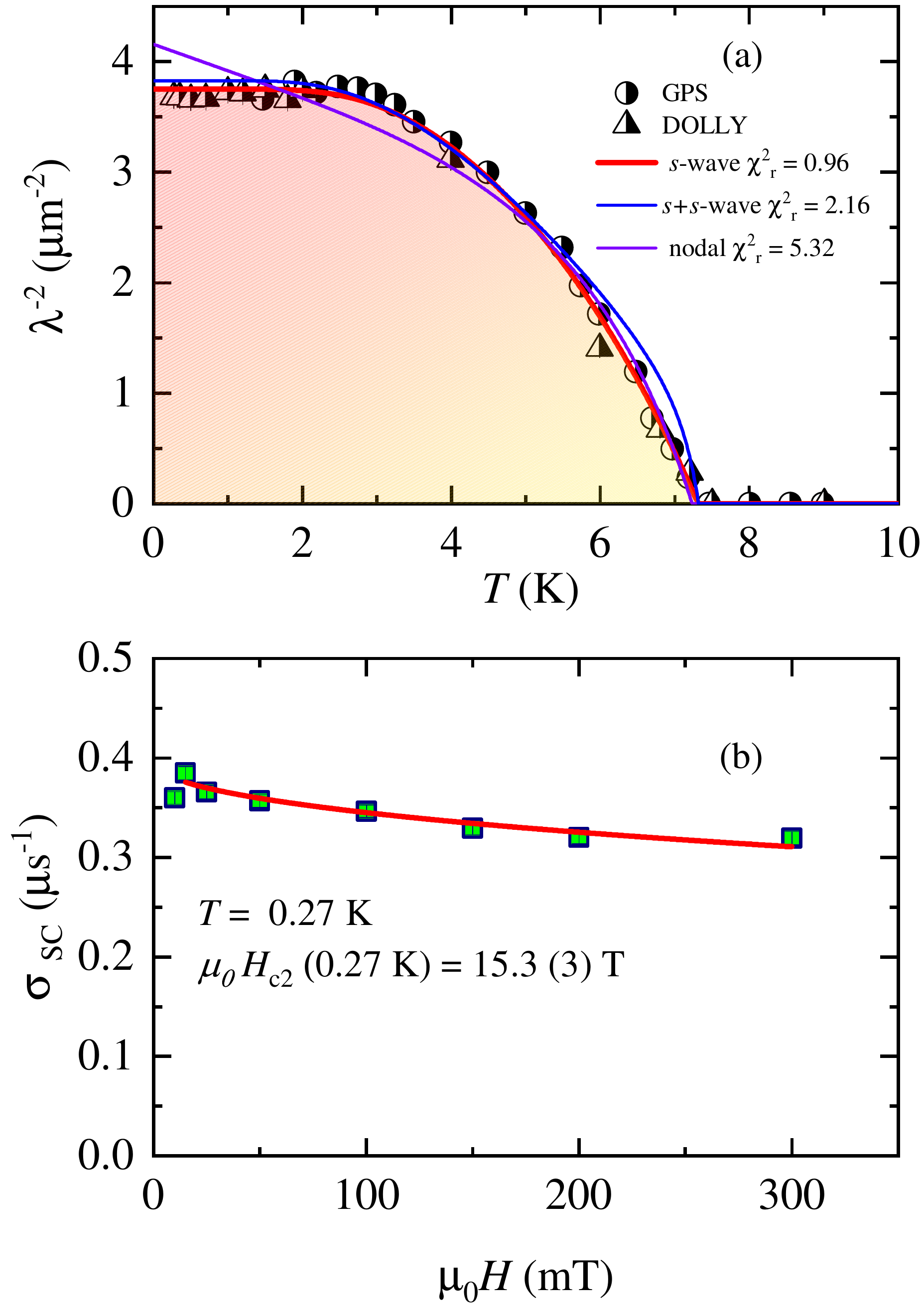}
\caption{(Color online) a) Temperature dependence of ${\lambda}^{-2}$ for NbIr$_2$B$_2$, measured in an applied field ${\mu}_{\rm 0}H=25$~mT. The solid lines correspond to different theoretical models as discussed in the text. (b) Field dependence of the superconducting muon spin depolarization rate at 0.25~K fitted with an isotropic single $s$-wave gap model (solid red line).}
\label{fig4}
\end{figure}

In the present case, as seen from the Figure~\ref{fig4}a, the experimentally obtained  ${\lambda}^{-2}(T)$ is best described by a momentum-independent single $s$-wave model with a gap value of $\Delta_0$ = 1.32(2)~meV and $T_{\rm c}$ = 7.3(2)~K. Thus obtained $T_{\rm c}$  is in good agreement with that estimated from other measurements [see Fig. 1(b)]. In our previous report\cite{Gornicka}, the specific heat data was fitted using two models: single gap $s$-wave and two gap $s+s$-wave. Interestingly, the gap value determined from $\mu$SR studies, matches very well with that obtained previously using a single gap (1.35(6)~meV) $s$-wave  model \cite{Gornicka}. Our attempt to model ${\lambda}^{-2}(T)$ using two gaps was unsuccessful as the relative weight of the gaps was reaching a value close to 0 implying single gap is more appropriate. Further, we also tried to fit the data fixing the gap values to those obtained in Ref \cite{Gornicka} (blue solid line in Fig~\ref{fig4}a). This also failed to model the experimentally observed temperature dependence of ${\lambda}^{-2}(T)$. We note that even though the  $s+s$-wave picture describes the heat capacity data slightly better \cite{Gornicka}, the question whether NbIr$_2$B$_2$ is a single gap $s$ wave or two gap $s+s$ wave superconductor remained speculative. Therefore, from the $\mu$SR experiments we confirm the presence of single superconducting gap.


To further corroborate a fully gaped state, we performed the field dependence of the TF-${\mu}$SR experiments. The TF-relaxation rate ${\sigma}_{\rm sc}(B)$ measured at 0.27~K is shown in Figure~\ref{fig4}b. During measurement, each point was obtained by field-cooling the sample from 12~K (above $T_{\rm c}$) to 0.27~K.  ${\sigma}_{\rm sc}$ first increases with increasing magnetic field until reaching a maximum at 15~mT followed by a continuous decrease up to the highest field (300~mT) investigated. Such field dependence resembles with the form expected for an single gap $s$-wave superconductor with an ideal triangular vortex lattice. Furthermore, the observed ${\sigma}_{\rm sc}(B)$ curve at fields above the maximum, can be analyzed using the Brandt formula (for an $s$-wave superconductor)~\cite{Brandt2},


\begin{multline}
\sigma_{\mathrm{sc}}~\left[\mu \mathrm{s}^{-1}\right]= 4.83\times 10^4\left(1-\frac{H}{H_{\rm {c2}}}\right)\\
\times\left[1+1.21\left(1-\sqrt{\frac{H}{H_{\rm c2}}}\right)^3\right]\lambda^{-2}~\left[\rm {nm}^{-2}\right].
\label{Field_dep}
\end{multline}

\noindent which provides an estimate of upper critical field at 0.27~K $\mu_0 H_{\rm c2}(0.27~K)=15.8(3)$~T which is in good agreement with the value obtained from the electrical resistivity measurements (see above).

\section{DISCUSSION}

The upper critical field- temperature phase diagram manifests a linear relationship (Figure~\ref{fig1}d). Such behavior is quite rare and strongly suggests unconventional superconductivity in NbIr$_2$B$_2$. It is worth noting that a nearly linear $\mu_0H_{\rm c2}(T)$ behavior was observed previously in two-band superconductors,  Lu$_2$Fe$_3$Si$_5$ \cite{Nakajima} and Ba(Fe$_{1-x}$Co$_x$)$_2$As$_2$ \cite{Hanisch}. However, Kogan and Prozorov showed that $H_{\rm c2}(T)$ linearity might be caused by competing effects of the equatorial nodes and of the Fermi surface anisotropy \cite{Kogan}.

Considering the ${\lambda}^{-2}(T)$ dependence observed for NbIr$_2$B$_2$, it is important to note that the ($p_x$ + $ip_y$) pairing symmetry is also characterized by the full gap and would also give saturated behavior of ${\lambda}^{-2}(T)$ at low temperatures. But, the possibility of $p_x$ + $ip_y$ pairing can be unequivocally excluded by the absence of TRS-breaking state. Further, to outwit any possibility of having nodes (see supplemental of Ref\cite{Gornicka}) in the superconducting gap structure, we also tested nodal gap symmetry (see Fig~\ref{fig4}a), which was found to be inconsistent with the data. Therefore, we ascertain a nodeless or fully gapped state as the most plausible superonducting pairing state in  NbIr$_2$B$_2$. This conclusion is also supported by the field dependence of depolarization rate.

The ratio of the superconducting gap to $T_{\rm c}$ was estimated to be (2$\Delta_0$/$k_B~T_{\rm c}$)$\sim$~4.2 which is consistent with the strong-coupling limit BCS superconductors \cite{GuguchiaNatcomm}. Interestingly, for the Bose-Einstein condensation (BEC) like picture, a similar ratio can also be expected. Thus, just from the ratio 2$\Delta_0$/$k_B~T_{\rm c}$, we cannot effectively distinguish between BCS or BEC condensations\cite{UemuraBEC}. In this regard, $T_{\rm c}/\lambda^{-2}(0)$ ratio is a quite crucial parameter to address this conjecture. Within the picture of BEC to BCS crossover \cite{Uemura,Uemura2}, systems exhibiting small $T_{\rm c}/\lambda^{-2}(0)$ are considered to be on the BCS-like side, whereas the large value of $T_{\rm c}/\lambda^{-2}(0)$ $\sim$1–20 and the linear relationship between $T_{\rm c}$  and $\lambda^{-2}(0)$ is expected only on the BEC-like side and is considered a hallmark feature of unconventional superconductivity\cite{GuguchiaNatcomm,Prando,GuguchiaFeAs,DasPdBi2}. For NbIr$_2$B$_2$, we obtained the ratio $T_{\rm c}\left[\rm K\right]/\lambda^{-2}(0) \left[\mu\rm{m}^{-2}\right] $ $\sim$ 2 which is intermediate between the values observed for electron-doped ($T_{\rm c}/\lambda^{-2}(0)\sim~1$) and hole-doped ($T_{\rm c}/\lambda^{-2}(0)\sim~4$) cuprates \cite{Uemura,Uemura2, Shengelaya}. This result yields strong evidence for an unconventional pairing mechanism in NbIr$_2$B$_2$ which also exhibits linear temperature dependency of the upper critical field.

In unconventional superconductors, one of the important parameters defining superconductivity is the superconducting carrier density, $n_s$. Within the London theory~\cite{Sonier}, the penetration depth is directly related to microscopic quantities such as the effective mass, $m^*$, and $n_{\rm s}$ via the relation $\lambda^2(0)$=$\left( m^*/\mu_0n_{\rm s}e^2\right)$. Here, $m^*$ can be estimated from the relation $m^*=(1+\lambda_{\rm e-ph})m_{\rm e}$ , where $\lambda_{\rm e-ph}$ is the electron–phonon coupling constant which was previously found out to be 0.74 from heat capacity \cite{Gornicka} and $m_{\rm e}$ is the electron rest mass. From $\lambda^{-2}(T)$ dependence, we determined $\lambda (0)$ = 516 (3)~nm. Thus, for NbIr$_2$B$_2$, we estimated $n_s = 1.88 \times 10^{26}$~m$^{-3}$. This value is comparable to that observed in other unconventional superconductors namely, ZrRuAs ($n_s = 2.1 \times 10^{26}$~m$^{-3}$)\cite{DasZrRuAs}, Nb$_{0.25}$Bi$_2$Se$_3$  ($n_{\rm s} = 0.25\times 10^{26}~$m$^{-3}$)~\cite{DasNbBiSe},  K$_2$Cr$_3$As$_3$ ($n_{\rm s} = 2.7\times 10^{27}~$m$^{-3}$)~\cite{Adroja} etc.  Thus, the relatively high value of $T_{\rm c}$ and low value of $n_{\rm s}$ in NbIr$_2$B$_2$ also suggestive of unconventional superconductivity in this compound. Previous band structure calculation \cite{Gornicka} shows existence of two Fermi-surface sheets in NbIr$_2$B$_2$. Therefore, Hall conductivity measurements are called for on this compound. This will be decisive to address the question whether the single-gap superconductivity in NbIr$_2$B$_2$, as seen through the microscopic experimental probe such as $\mu$SR, originates from the superconducting gap occurring only on an electron or hole-like Fermi surface.

\section{Conclusion}
In conclusion, we provide a detailed microscopic understanding of the superconducting gap structure of novel noncentrosymmetric superconductor NbIr$_2$B$_2$ . The temperature as well as the field dependence of $\lambda^{-2}$ was investigated by means of $\mu$SR experiments which allowed us to determine the zero-temperature magnetic penetration depth $\lambda (0)$. Interestingly, the $T_{\rm c}/\lambda^{-2}(0)$
ratio is comparable to those of high-temperature unconventional superconductors, signaling unconventional nature
of superconductivity in this compound. The linear temperature dependence of upper critical field also supports this idea. Furthermore, $\mu$SR, which is an extremely sensitive magnetic probe, reveals absence of any spontaneous magnetic fields that would be expected for a TRS-breaking state in the bulk of  NbIr$_2$B$_2$. Therefore, our results classify NbIr$_2$B$_2$ as an unconventional timereversal invariant and single gap superconductor.

\begin{acknowledgments}
The muon spectroscopy studies were performed at the Swiss Muon Source (S${\mu}$S) Paul Scherrer Insitute, Villigen, Switzerland. DD and ZG thank C. Baines for the technical assistance during DOLLY experiments. A portion of this work was performed at the National High Magnetic Field Laboratory, which is supported by National Science Foundation Cooperative Agreement No. DMR-1644779 and the State of Florida. The research performed at the Gdansk Tech. was supported by the National Science Centre (Poland) grant (UMO-2018/30/M/ST5/00773).
\end{acknowledgments}

\end{document}